\documentclass[conference]{IEEEtran}
\IEEEoverridecommandlockouts
\usepackage{cite}
\usepackage{amsmath,amssymb,amsfonts}
\usepackage{algorithmic}
\usepackage{graphicx}
\usepackage{textcomp}
\usepackage{xcolor}
\def\BibTeX{{\rm B\kern-.05em{\sc i\kern-.025em b}\kern-.08em
    T\kern-.1667em\lower.7ex\hbox{E}\kern-.125emX}}
\begin{document}

\title{CovXR: Automated Detection of COVID-19 Pneumonia in Chest X-Rays through Machine Learning\\}

\author{%
  \makebox[.5\linewidth]{Vishal Shenoy}\\ \textit{Cupertino High School}\\ Cupertino, CA \\ vishal.nshenoy@gmail.com\\
    \and \makebox[.5\linewidth]{Sachin Malik}\\ \textit{Stanford University Department of Radiology}\\ Stanford, CA \\ sbmalik@stanford.edu\\
}

\maketitle

\begin{abstract}
Coronavirus disease 2019 (COVID-19) is the highly contagious illness caused by severe acute respiratory syndrome coronavirus 2 (SARS-CoV-2). The standard diagnostic testing procedure for COVID-19 is testing a nasopharyngeal swab for SARS-CoV-2 nucleic acid using a real-time polymerase chain reaction (PCR), which can take multiple days to provide a diagnosis. Another widespread form of testing is rapid antigen testing, which has a low sensitivity compared to PCR, but is favored for its quick diagnosis time of usually 15-30 minutes. Patients who test positive for COVID-19 demonstrate diffuse alveolar damage in 87\% of cases. Machine learning has proven to have advantages in image classification problems with radiology. In this work, we introduce CovXR as a machine learning model designed to detect COVID-19 pneumonia in chest X-rays (CXR). CovXR is a convolutional neural network (CNN) trained on over 4,300 chest X-rays. The performance of the model is measured through accuracy, F1 score, sensitivity, and specificity. The model achieves an accuracy of 95.5\% and an F1 score of 0.954. The sensitivity is 93.5\% and specificity is 97.5\%. With accuracy above 95\% and F1 score above 0.95, CovXR is highly accurate in predicting COVID-19 pneumonia on CXRs. The model achieves better accuracy than prior work and uses a unique approach to identify COVID-19 pneumonia. CovXR is highly accurate in identifying COVID-19 on CXRs of patients with a PCR confirmed positive diagnosis and provides much faster results than PCR tests. 
\end{abstract}


\section{Introduction}
The coronavirus disease 2019 (COVID-19) began in late 2019 after the diagnosis of an unknown viral pneumonia in Wuhan, China [1]. COVID-19 is the highly contagious illness caused by severe acute respiratory syndrome coronavirus 2 (SARS-CoV-2), and has spread across the globe through a devastating pandemic. As of July 2021, there have been over 190 million confirmed cases of COVID-19 and over 4 million deaths caused by COVID-19 globally [2]. 

Patients with COVID-19 may exhibit flu-like symptoms including fever, cough, chills, fatigue, muscle aches, loss of taste/smell, headache, and congestion [3]. Severe symptoms of COVID-19 include difficulty breathing, pressure in the chest, and new confusion. Up to 14\% of patients have been reported to develop severe symptoms which require hospitalization [4, 5].

The standard diagnostic testing procedure for COVID-19 examines a nasopharyngeal swab for SARS-CoV-2 nucleic acid using a real-time polymerase chain reaction (PCR). PCR tests have been validated by the US Food and Drug Administration (FDA) as an effective method of diagnosis. The sensitivity of most FDA-approved commercial PCR tests is dependent on many factors that include the quality of the specimen and time from exposure. PCR tests typically require one to three days to result and assist in making a diagnosis for the patient [4].

COVID-19 diagnostic testing can also be done with antigen tests, which typically provide results in as little as  15 minutes [6]. This speed comes at a slight disadvantage: some studies suggest that the sensitivity of antigen testing is only 80\%. However, the specificities for both PCR tests and antigen tests are nearly 100\% [4, 7]. 

\begin{figure}[htbp]
\centerline{\includegraphics{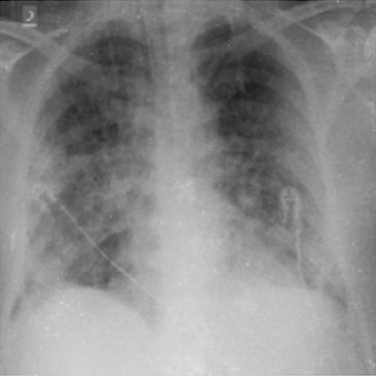}}
\caption{Chest X-ray of patient with COVID-19 pneumonia.}
\label{fig}
\end{figure}

Patients who test positive for COVID-19 demonstrate diffuse alveolar damage in 87\% of cases [4]. These findings are  especially visible on chest x-rays (CXR) in patients with severe COVID-19 symptoms (see Figure 1). A CXR can be generated in under 15 minutes and has several advantages over COVID-19 PCR and antigen testing methods [8]. Generally, radiology equipment is readily accessible in even the most resource-limited healthcare settings in the United States. Equipment can be operated in an isolated environment and easily disinfected after use.  

The most common views for CXRs are posteroanterior and anteroposterior. In posteroanterior (PA) views, the x-ray beam enters through the back and exits out the front of the chest. In anteroposterior (AP) views, the x-ray beam enters through the front and exits out the back of the chest. AP CXRs are harder to read than PA CXRs and are used for situations where it is difficult to obtain an AP CXR, such as when the patient is bedridden. 

Machine learning has proven to have advantages in image classification problems. Over the past decade, the usage of machine learning in radiology has increased tremendously and shown high performance through many applications [10]. We present CovXR as a machine learning algorithm to detect COVID-19 pneumonia in CXRs. This paper describes the creation of CovXR and analyzes the performance of the model.

\section{Prior Work}
Some academic groups have attempted to solve the challenge of detecting COVID-19 pneumonia in chest X-rays using machine learning. CovXR is similar to two key pieces of research that we summarize below.

Wehbe et al. constructed a model for detecting COVID-19 pneumonia with transfer learning and an ensemble of convolutional neural networks (CNN) [11]. The model was trained and validated on 14,788 images with 4,253 COVID-19 positive CXRs. Their approach obtained a validation accuracy of 82\%, sensitivity of 71\%, and specificity of 92\%. Another paper by Zhang et al. trained a model on CXRs with COVID-19 pneumonia and CXRs with non-COVID-19 pneumonia [12]. This approach achieved a peak sensitivity of 89\% and a peak specificity of 90\%. 

In our opinion, while the specificity achieved by Wehbe et al. is impressive, the sensitivity is outperformed by many trained radiologists [11]. These metrics may also be insufficient due to severe class imbalance in the training set. The model created by Zhang et al. approaches the problem of differentiating between types of pneumonia. This work achieves impressive metrics, but requires more effort to deploy than simply inputting a CXR through a model: a skilled radiologist would have to make a judgement on whether the CXR exhibits damage features, which requires time and expertise. According to our literature review, other approaches use different forms of image preprocessing and achieve lower specificity/sensitivity metrics. We believe that CovXR is the most competent and robust algorithm for detecting COVID-19 pneumonia to-date.  

\section{Methods}

Advances in hardware and software have allowed for rapid progress in machine learning systems which learn from data to predict labels for new inputs. These systems require minimal human involvement and can achieve great accuracy with their label predictions. Machine learning is the process of training a classification algorithm on a dataset, resulting in a model which can autonomously identify patterns and make decisions. In traditional programming methods, data is processed according to rules manually input by humans. By contrast, in machine learning, data is processed according to rules generated by a software system.  

CovXR leverages a machine learning technique called transfer learning. Transfer learning is a method where a pre-trained model is reused for a new task. These pre-trained models are trained on a large dataset and contain multiple convolutional layers. The datasets contain images which are usually unrelated to the transfer learning task. However, since they reuse the weights of a complex model, transfer learning models are able to achieve high accuracy with limited amounts of data [13]. This approach offers a better starting point for machine learning tasks and thus requires a shorter amount of time to train a model [14]. Figure 2 summarizes four steps in the development of our model. 

\begin{figure}[htbp]
\centerline{\includegraphics{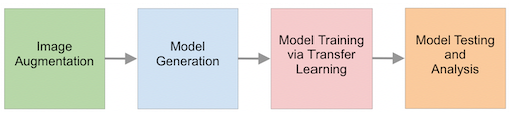}}
\caption{Development pipeline for CovXR.}
\label{fig}
\end{figure}

A pre-trained ResNet50 model [15] was leveraged for the creation of CovXR. ResNet50 is a CNN with 48 convolutional layers that has been trained on over one million images from the ImageNet database [16]. ResNet, which is short for Residual Network, is used as the backbone for many machine learning tasks and has achieved high accuracy with various computer vision problems.

A dataset of 16,351 CXRs was collected from Kaggle [17, 18]. Kaggle is an online repository of community published data. The dataset implemented in this project was constructed from a wide collection of publicly accessible CXR data sources. The dataset is provided in two parts: a training set with 15,951 scans and a testing set with 400 scans. The training set contains 13,793 COVID-19 negative scans and 2,158 COVID-19 positive scans. The testing set contains 200 COVID-19 negative scans and 200 COVID-19 positive scans. The training set contains CXRs from 14,978 unique patients and the testing set contains CXRs from 378 unique patients. Both sets contain a mix of PA and AP CXRs.

The training dataset is severely unbalanced: there are over six times as many COVID-19 negative scans than COVID-19 positive scans. Imbalanced datasets are a challenge for machine learning models because models are designed to assume an equal number of examples per classification label [19]. An extreme imbalance results in poor predictive performance, especially when predicting labels for the minority class [19]. Thus, it was necessary to balance the dataset with equal numbers of positive and negative scans before constructing a machine learning algorithm. A total of 2,158 COVID-19 negative scans and an equal number of COVID-19 positive scans were used to train CovXR. The training dataset was divided into an 80\% train and 20\% validation split.

\begin{figure}[htbp]
\centerline{\includegraphics{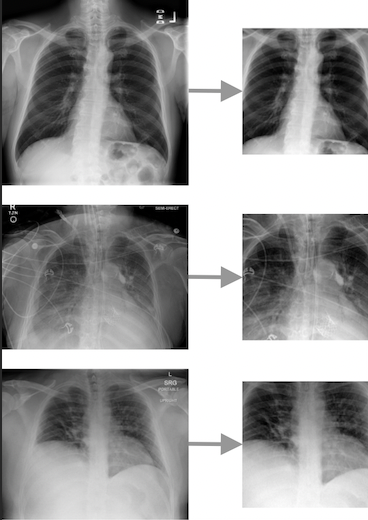}}
\caption{Three images before and after being augmented with a colorspace transition, 30\% zoom, and compression to 224 by 224 pixels.}
\label{fig}
\end{figure}

The training images were augmented using the Keras framework. To fit the ResNet50 architecture, training images were converted from the RGB color space to the BGR color space and resized to 224 by 224 pixels. The color space transform reverses the color channels to match required specifications. Training images were zoomed in by a factor of 30\% to provide a clear visualization of the lungs, cropping out the neck, shoulders, and abdomen from the CXR scan. This process enabled the model to recognize important features within the lung instead of unnecessary patterns in other areas of the body. In addition, some training images received a vertical flip to increase the diversity of the training data. These image altercations are automatically performed on inputs during training. Figure 3 illustrates these transformations on real CXRs. 

CovXR was programmed in a Jupyter Notebook using the Python 3 language [20, 21]. The Tensorflow 2.5 and Keras packages were utilized when building and training the transfer learning model [22, 23]. Adam was used as an optimization algorithm for stochastic gradient descent [24]. The Nvidia GeForce RTX 2060 graphics processing unit (GPU) was used as the hardware for training CovXR [25]. The model was trained across ten epochs with accuracy, precision, and recall metrics being measured. Figure 4 breaks down the model’s layers, which include a pooling operation layer to reduce the dimensions of feature maps, two densely-connected neural network layers to receive inputs from previous layers, a normalization layer to standardize inputs in each batch of 64 images, and a dropout layer to prevent overfitting. All layers, except for the last, were connected via rectified linear units (ReLu). A sigmoid activation function was used to connect the final dense layer in order to map output probabilities between 0 and 1. We selected binary cross-entropy as our loss function for optimizing our network parameters (see Equation \ref{binary}, \emph{y} = binary indicator that observation is COVID-19 negative or positive, \emph{p} = predicted probability that observation is COVID-19 positive). 

\begin{figure}[htbp]
\centerline{\includegraphics{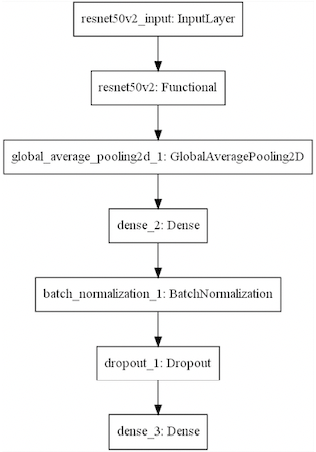}}
\caption{Summary of CovXR model layers.}
\label{fig}
\end{figure}

\begin{equation*}\label{binary}
H_{binary}(p) = 
-(ylog(p) + (1-y)log(1-p))
\tag{1}
\end{equation*}

Cross-entropy loss (also known as logarithmic loss) measures the performance of a predictive model whose output is a probability value between 0 and 1, where 1 is indicative of COVID-19 symptoms within an input CXR. This loss function increases as the predicted probability of a sample diverges from the true class label. Properly classified inputs contribute a value of 0 to the cummulative loss.

\section{Results}

The model was validated through the calculation of several performance metrics on an entirely new testing set of 400 images. Sensitivity (also known as recall, see Equation 2) measures CovXR’s ability to correctly identify patients with COVID-19. Specificity (also known as precision, see Equation 3) measures CovXR’s ability to correctly identify patients without COVID-19. The F1 score (see Equation 4) is defined as the harmonic mean of sensitivity and specificity; it is proven to be a good measure of competence in machine learning models [26]. Accuracy is the percentage of correct predictions across all cases with and without COVID-19 (see Equation 5). CovXR achieved an accuracy of 95.5\% and an F1 score of 0.954 on the testing set of images. Sensitivity is 93.5\% and specificity is 97.5\%. A confusion matrix for the testing set used to gain these metrics is shown in Figure 5. 

\begin{figure}[htbp]
\centerline{\includegraphics{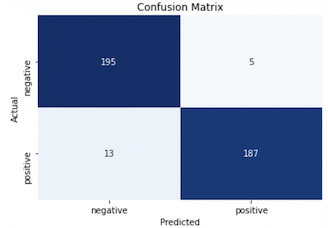}}
\caption{Confusion matrix illustrating predictions on 400 test images passed through CovXR.}
\label{fig}
\end{figure}

\begin{equation*}
sensitivity = 
\frac{true \enspace positive}{all \enspace positive}
\tag{2}
\end{equation*}

\begin{equation*}
specificity = 
\frac{true \enspace negative}{all \enspace negative}
\tag{3}
\end{equation*}

\begin{equation*}
F1 \enspace score = 2 \cdot
\frac{sensitivity \cdot specificity}{sensitivity+specificity}
\tag{4}
\end{equation*}

\begin{equation*}
accuracy = 
\frac{all \enspace correct}{all \enspace samples}
\tag{5}
\end{equation*}
\\

\begin{figure}[htbp]
\centerline{\includegraphics{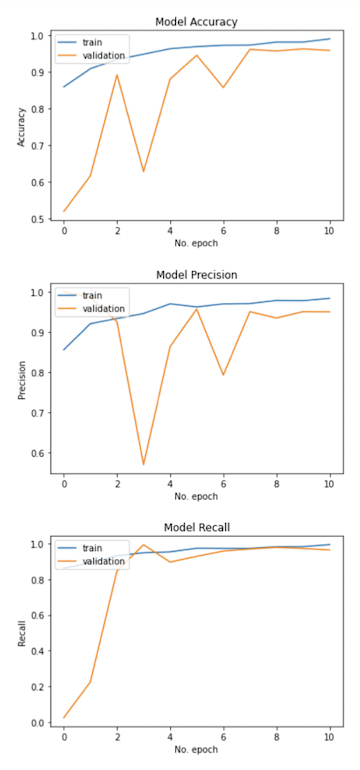}}
\caption{Graphs showing training versus validation metrics for accuracy, precision, and recall.}
\label{fig}
\end{figure}

The higher specificity compared to sensitivity indicates that CovXR is better at correctly identifying healthy patients without COVID-19. The algorithm detected more false negatives than false positives, and while this is typically an area of concern in medical implementations, the difference between sensitivity and specificity is very low, at  less than 4\%. However, both metrics are very high, suggesting that the system is still highly competent in detecting patients with and without COVID-19.

Training metrics show how well a model is performing on the data it is trained with. Validation metrics show how well a model is performing on new data it has never seen before. Overfitting happens when a machine learning algorithm learns specific rules for the training set which do not apply beyond the set. Figure 6 plots the training and validation metrics for accuracy, precision, and recall over ten epochs. The closeness in training and validation metrics in later epochs prove that overfitting is essentially nonexistent.

\begin{figure}[htbp]
\centerline{\includegraphics{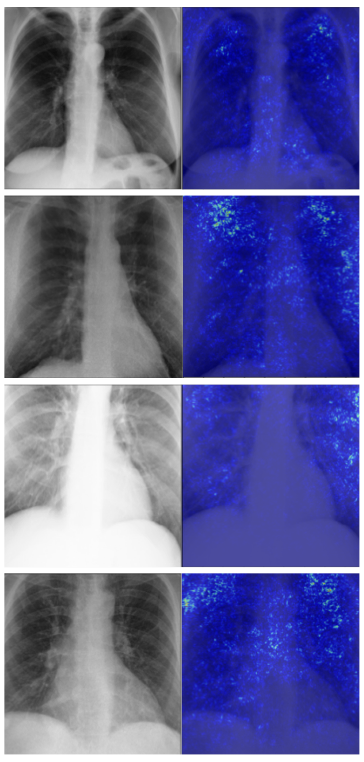}}
\caption{From top to bottom: saliency maps for true negative, false negative, false positive, and true positive chest X-ray inputs}
\label{fig}
\end{figure}

When analyzing a machine learning model, it is important to understand why the algorithm works. Saliency maps are a way to visualize the inner workings of machine learning models in the form of heats maps which highlight features within an image that the model is focused on [27]. Saliency maps were generated on a couple of images in the testing set to ensure that CovXR is identifying regions of interest within the lung. The maps in Figure 7 (see next page) confirm that the attention of our model is drawn to the proper regions in the images. 

\section{Conclusion}

In this paper, we presented CovXR, a machine learning algorithm which predicts COVID-19 pneumonia with high accuracy on CXRs of patients who are PCR positive; accuracy was above 95\% at 95.5\% and F1 score was above 0.95 at 0.954. Compared to the 80\% sensitivity of antigen testing, CovXR is much more accurate in diagnosing patients with COVID-19 [7]. These metrics are also higher than those achieved by previous machine learning approaches to identifying COVID-19 pneumonia [11, 12]. CovXR is as quick as COVID-19 antigen testing and has the unique advantage of accessibility: radiology equipment is available in almost all hospital and clinical environments in the United States. In comparison to COVID-19 PCR assays, CovXR trades a slight decrease in accuracy for a tremendously faster diagnosis. A quick diagnosis is important because COVID-19 positive patients can self-isolate or quarantine sooner to help stop the spread of the virus.

CovXR identifies lung abnormalities which may be shared with diseases other than COVID-19. Many other lung infections have overlapping abnormalities with COVID-19, which is why patient history is important when reaching a diagnosis with CovXR. Also, CXRs are not sensitive in the early stages of disease and are therefore not used to "rule out" disease in a patient. CovXR maintains advantages of an accuracy similar to experienced radiologists [11] and rapid results that can help with triaging patients in resource limited environments. In practice, CovXR can help doctors by providing a second opinion on COVID-19 diagnosis and highlighting a feature map for areas of interest.

\section{Future Work}
There are many ways to improve CovXR. We would like to augment CovXR to implement an ensemble transfer learning model; this would consider the decisions of multiple transfer models in a final COVID-19 positive or negative diagnosis, leading to a higher accuracy system. We would also like to research ways of easily implementing this technology in real hospital and clinical settings. This could potentially be done through a CovXR mobile application or website.

\end{document}